\documentclass[preprint,11pt]{JHEP3} % 10pt is ignored!

\JHEPspecialurl{http://jhep.sissa.it/JOURNAL/JHEP3.tar.gz}

\usepackage{epsfig,multicol,amsmath}

\def\beq{\begin{equation}}
\def\eeq{\end{equation}}
\def\bea{\begin{eqnarray}}
\def\eea{\end{eqnarray}}

\def\eq#1{{Eq.~(\ref{#1})}}
\def\fig#1{{Fig.~\ref{#1}}}
\newcommand{\bas}{\bar{\alpha}_S}
\newcommand{\as}{\alpha_S}

\newcommand{\Lb}{\left(}
\newcommand{\Rb}{\right)}

\parskip=\itemsep               %?

\def\thefootnote{\fnsymbol{footnote}}

\title{ A geometrical scaling  solution to  the BFKL Pomeron Calculus in the saturation domain}
\author{\large E. Levin\thanks{Email: leving@post.tau.ac.il,
levin@mail.desy.de, elevin@quark.phy.bnl.gov;} \\
Department of Particle Physics, School of Physics and Astronomy\\ Raymond and Beverly Sackler Faculty
of Exact Science\\  Tel Aviv University, Tel Aviv, 69978, Israel}

\abstract{ In this paper an analytical solution is found for the BFKL Pomeron calculus in QCD,
in which all Pomeron loops have been included. This solution manifests the geometrical scaling
behaviour and matches with the solution to the linear evolution equation on the line $r^2= 1/Q^2_s(x)$,
 where $r$ is the dipole size and $Q_s$ is the saturation momentum.}

 \keywords{BFKL Pomeron,  Generating functional, Fokker-Planck equation, Exact solution }

\preprint{  TAUP -2830-06\\
\today}
\begin{document}

\def\thefootnote{\arabic{footnote}}
\section{Introduction}
\label{sec:Int}
The BFKL Pomeron calculus\cite{GLR,MUQI} gives the simplest and the most
transparent  approach to the high energy amplitude in QCD. This approach is based on the BFKL Pomeron
\cite{BFKL} and the  interactions between the BFKL Pomerons \cite{BART,BRN,NP,BLV}, which are
taken into account in the spirit of the  Gribov Reggeon Calculus \cite{GRC}. It can be
written in the form of  the functional integral (see Ref. \cite{BRN})  and  formulated in
terms  of the generating functional (see Refs. \cite{MUCD,L1,L2,L3,L4}).

In Ref. \cite{KLP} \footnote{See Eq.(3.18), Eq.(3.55) , Eq.(4.97) and Eq.(4.98)  of Ref. \cite{KLP} for more details.}
(see also Ref. \cite{MSX}) it is shown that the BFKL Pomeron calculus can be reduced to the following
functional integral
\beq \label{FI}
Z[\Phi,\Phi^+]\,\,=\,\,\int\,
\,D\,\Phi(b,k;Y)\,D\,\Phi^+(b,k;Y)\,e^{S[\Phi,\Phi^+]}
\eeq
with
\beq  \label{FIMR}
S\,\,=\,\,S_0\,+\,S_I\,\,=
\,\,\int\,d^2 \,b\,d^2\,k\,\,\Phi^+\left(b, k;
Y\right)\,\,
\left( \frac{\partial \Phi\left(b,k;Y\right)}{\partial\,Y}\,\,-\,\,\
 \frac{\bas}{2\pi}\,\left(\int\,d^2\,k'\,K\left(k,k'\right)\,
\Phi\left(b,k';Y\right)\,\,+ \right.\right.
\eeq
$$
\left. \left.
\,\, \Phi^+(b,k;Y)\,\Phi(b,k;Y) \,\,\,
+\,\,\,\as^2 \,\left\{\Phi\left(b,k;Y\right)\,\Phi^+\left(b,k;Y\right)\,\,-\,\,\Phi^+\left(b,k;Y\right)\,\Phi\left(b,k;Y\right)\,\Phi\left(b,k;Y
\right)
\right\}\,\right) \right)
$$
where $K\left(k,k'\right)$ is the BFKL kernel in the momentum representation, namely,
\beq \label{BFKLKR}
\int\,d^2\,k'\,\,K\left(k,k'\right)\,\,\Phi\left(b,k';Y\right)\,\,=\,\,
 \int\,d^2\,k'\,\,\frac{\Phi\left(b,k';Y\right)}{(k -
k')^2}\,\,-\,\,k^2\,\int\,d^2\,k'\,\frac{\Phi\left(b,k;Y\right)}{(k - k')^2\,(k'^2 \,+\,(k - k')^2)}
\eeq

The theory determined  by the functional integral of \eq{FI},  can be written \cite{KLP}  in the equivalent form introducing the generating functional  $Z(Y,[u(b,k)])$ defined as \cite{MUCD}
\beq \label{ZMR}
Z\left(Y\,-\,Y_0;\,[u(b_i,k_i)] \right)\,\,\equiv\,\,
\eeq
$$
\equiv\,\,\sum_{n=1}\,\int\,\,
P_n\left(Y\,-\,Y_0;\,b_1, k_1; \dots ; b_i, k_i; \dots ;b_n, k_n
 \right) \,\,
\prod^{n}_{i=1}\,u(b_i, k_i) \,d^2\,x_i\,d^2\,k_i
$$
where $u(b_i,k_i)$ are  arbitrary functions and $P_n\left(Y\,-\,Y_0;\,b_1, k_1; \dots ; b_i, k_i; \dots ;b_n, k_n \right)$ is the probability to find $n$-dipoles with transverse momenta $k_i$.

For this functional we have the following evolution equation \cite{KLP}
\begin{equation}\label{ZEQMR}
\frac{\partial \,Z\Lb Y-Y_0; [\,u(b,k)\,]\Rb}{
\partial \,Y}\,\,= \,\,\chi\,[\,u(b,k)\,]\,\,Z\Lb Y- Y_0; [\,u(b,k)\,] \Rb
\end{equation}
with
\begin{eqnarray}
\chi[u]\,\,&=&\,\, \,\int\,d^2\,b\, d^2\, k\, \left( \frac{\bas}{2 \pi}
 \left(- \int\,d^2\,k'\, K(k,k') u(b,k')\,\frac{\delta}{\delta u(b,k')} \,+\,u(b,k) \,u(b,k)\,
\,\frac{\delta}{\delta u(b,k)} \right)\,- \right. \label{chimr} \\
 & &\left. - \as^2
\left( u(b,k)  \,u(b,k) \,-\, u(b,k) \right) \,\,\frac{1}{2} \,\frac{\delta^2}{\delta
u(b,k)\,\delta
u(b,k)} \right)\,;
\label{VE21MR}
\end{eqnarray}

In this paper our goal is to solve \eq{ZEQMR} deeply in the saturation region or, in other words, for the dipoles which
sizes are much larger than the typical saturation size $1/Q_s(x)$.  In the next section we set the problem and will discuss the main assumptions that simplify the problem.
In section 3
 we discuss the mean field approximation. In this approximation the expression for $\chi[u]$ is simple and it  is determined by \eq{chimr}.
Solution in this approximation is known (see Ref. \cite{LT}) but we will solve this equation using a different method. Therefore, we view this section as a training ground for our method of finding a solution.

In section 4 we focus our efforts on searching a solution to the general equation (see \eq{ZEQMR}). We consider the solution which shows the geometrical scaling behaviour in the saturation domain and which matches with the linear evolution equation at $ r\, \approx\, 1/Q_s(x)$ where $ r $ is the dipole size and $Q_s(x)$ is the new scale:  saturation momentum.

In conclusions we summarize the main results of the paper and compare them  with other attempts to solve this equation .

\section{Problem setting and the main assumptions.}
The problem that we would like to solve, is the scattering of very small colourless dipole , which size is $r$,  with   the large target (let say with a nucleus with radius $R$, $R\,\gg\,r$).  We are interested in  finding the asymptotic behaviour of this amplitude deeply in the saturation region where $\as \ln(r^2\,Q^2_s)\,\,\gg\,\,1$.

Generally speaking , we need to fix initial and boundary conditions to solve  \eq{ZEQMR}.  These conditions are well known and they have the form

\bea
\mbox{Initial conditions:}\,\,\,\,\,\,\,\,\,& \,\,\,\mbox{at } Y=0 \,\,\,&\,\,\, Z\left(Y=0 ;[ u(b,k)] \right)\,\,= u(b,k)\,\,;\label{IC}\\
\mbox{Boundary conditions:}& \mbox{at } u_i=1 & \,\,\,\,\,Z\left(Y ;[ u(b,k)=1] \right)\,\,\,\,\,= \,\,\,\,\,\,\,\,1\,\,;\label{BC}
\eea
\eq{IC} means that at  low energy the interaction can be reduced to the  single  BFKL Pomeron exchange while \eq{BC}
follows directly from the meaning of $P_n$ as probability to find $n$-dipoles and reflects the normalization for the probability ($\sum P_n =1$).

  In terms of anomalous dimension $\gamma$ we replace the general BFKL kernel (see solid line in \fig{ker}
\beq \label{BFKLKER}
  \int \,K(k,k')\,e^{ (\gamma - 1)\,\ln(k'^2/k^2)}\,\,=\,\, \chi \Lb\gamma \Rb \,\,=\,\,2\,\psi(1) \,\,-\,\,\psi(\gamma) \,\,-\,\,\psi(1 - \gamma)
\eeq
by
\beq \label{CHIM}
\chi(\gamma)\,\,=\,\,\left\{\begin{array}{c}
\,\,\,\,\,\,\frac{1}{\gamma}\,\,\hspace*{1cm}\mbox{for}\,\,r^2\,Q^2_s\,\ll\,1\,; \\
\,\frac{1}{1\,-\,\gamma}\,\,\hspace*{1cm} \mbox{for}\,\,r^2\,Q^2_s\,\gg\,1\,;
\end{array} \right.
\eeq
(see dotted-dashed line in \fig{ker})

This approach is quite different from the diffusion approximation for  $\chi \Lb\gamma \Rb$ (see dotted
in \fig{ker}
\beq \label{bfkldif}
 \chi \Lb\gamma \Rb \,\,=\,\,2 \ln2\,\,+\,\,28\,\zeta(3)\,( \gamma - \frac{1}{2})^2
\eeq
  which was used in  Refs. \cite{MSX,STPH,EGM,IMS} to reduce the general equation to so called
Fisher-Kolmogorov-Pertrovsky-Piscounov equation\cite{FKPP} (F-KPP equation). $\zeta(x)$ in \eq{bfkldif} is the Riman zeta function.

The kernel of \eq{CHIM} sums the contributions of the order of $ \left(\as\,\ln(r^2\,\Lambda^2)\right)^n$
for $r^2\,Q^2_s\,\ll\,1$ where $\as\,\ln(r^2\,\Lambda^2)\,\,\gg\,\,1$; and it  leads to summation of the terms of the order
of $ \left(\as\,\ln(r^2\,Q^2_s)\right)^n$ in the kinematic region where $r^2\,Q^2_s\,\gg\,1$ and $as\,\ln(r^2\,Q^2_s)
\,\gg\,1$ \cite{MULG,LT}.

\FIGURE[h]{
\begin{minipage}{75mm}{
\centerline{\epsfig{file= 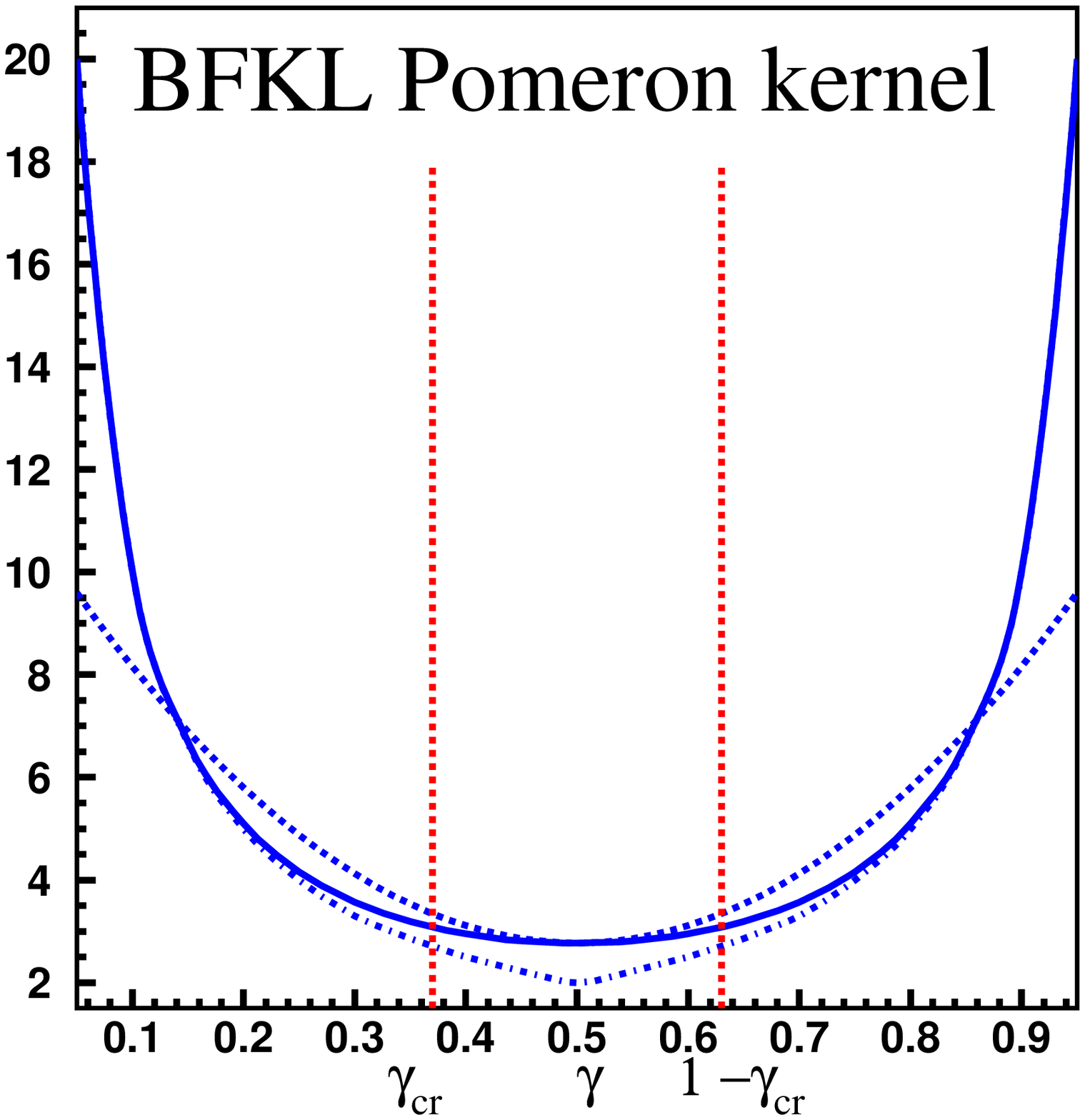,width=70mm}}
}
\end{minipage}
\caption{The kernel of the BFKL equation: the  solid line shows the exact expression of \protect\eq{BFKLKER};
the kernel of \protect\eq{CHIM} is plotted in dotted-dashed line; and the diffusion approximation of \protect\eq{bfkldif} for the BFKL  kernel, that leads to F-KPP equation, is presented by dotted line. $\gamma_{cr}$ is calculated using \protect\eq{QSCOR}.}
\label{ker}
}

  It means that this kernel can describe the behaviour of our systen deeply in the saturation region ($r^2\,Q^2_s\,\gg\,1$) or/and in the perturbative QCD region at $r^2\,Q^2_s \,\ll\,1$ while we cannot expect that
our approximation will be able to reproduce the exact solution in the vicinity of the saturation scale $r^2\,Q^2_s \,\approx\,1$). These features of this kernel one can see from \fig{ker}. \fig{ker} shows that the approximate kernel of \eq{CHIM}  can describe the exact BFKL kernel only at $\gamma$ close to
0 and 1 while  it leads to a contribution which  considerably differs from the exact kernel at $\gamma = 1/2$.
For  $\gamma$ close to $1/2$ we can use the diffusion approximation of \eq{bfkldif} which fails to reproduce the correct behaviour inside the saturation region.  One can see this since  we cannot  obtain  the correct asymptotic behaviour \cite{LT} in the mean field approximation with the kernel of \eq{bfkldif}.  On the other hand, the kernel of \eq{CHIM} cannot describe the behaviour of the scattering amplitude in the region close to the saturation scale. In particular, \eq{CHIM}
gives the saturation scale as
\beq \label{QSCHIM}
\ln(Q^2_s\,R^2)\,\,=\,\,4 \bas \,Y\,\,=\,\,  4 \bas \,\ln(1/x)
\eeq
where $\bas = (N_c/\pi)\,\as$, while the correct saturation scale is equal \cite{GLR,MUTR,MP}
\beq \label{QSCOR}
\ln(Q^2_s\,R^2)\,\,=
\,\,\frac{\chi\Lb \gamma_{cr}\Rb}{ 1 -  \gamma_{cr}}\,\,\,\,\,\mbox{with} \,\,\gamma_{cr} \,\,\mbox{determined by equation:}\,\,\,\,
\frac{\chi\Lb \gamma_{cr}\Rb}{ 1 -  \gamma_{cr}}\,\,=\,\,-\,\frac{d \chi\Lb \gamma_{cr}\Rb}{d
\,\gamma_{cr}}
\eeq
\eq{QSCOR} gives $\gamma_{cr} = 1/2$  that reduces \eq{QSCOR} to \eq{QSCHIM} while the correct value of $\gamma_{cr}$ is $\gamma_{cr} \,=\,0.37$ (see \fig{ker}).

We used two additional assumptions to reduce the general functional integral for the BFKL Pomeron calculus given in Ref. \cite{BRN} to the simple form of \eq{FIMR}. Namely, we assume  that at high energy we can use the BFKL Pomeron at $\gamma $ close to $1/2$ which leads to
~
$$
\left( L_{12}\,\, \equiv \,\,x^4_{12} \nabla^2_{x_1}\,\,\nabla^2_{x_2} \right)\,\,
 G\left(x_1,x_2;Y| x'_1,x'_2;Y'\right)\,\,=
 $$
\beq \label{L12}
 =\,\,G\left(x_1,x_2;Y| x'_1,x'_2;Y'\right)
\eeq
where $G\left(x_1,x_2;Y| x'_1,x'_2;Y'\right)$ is the Pomeron Green's function.

This  assumption works quite well both for diffusion approximation for the BFKL kernel (see \eq{bfkldif}) and for the kernel of  \eq{CHIM}
since $\gamma_{cr}$ for this kernel is equal to $1/2$ ($\gamma_{cr} = 1/2$).

It should be mentioned that \eq{FIMR} is written assuming
   that
\beq \label{B}
b\,\,\equiv\,\,\frac{x_1 \,+\,x_2}{2}\,\,\gg\,\,x_1 \,-\,x_2\,\,;
\eeq
~
Indeed, this assumption looks natural for the mean field approximation,since
 we expect that the typical size of the dipoles will be of the order of $1/Q_s(x)$ ($Q_s(x) $ is the
saturation scale) while the typical impact parameter of the scattering dipole should be much larger
(at least
of the order of
the size of  a  target ($R$) ). However, in the Pomeron loop we have integration over $b$  and  we should study the $b$ dependence more carefully.  Generally speaking, the integration over impact parameter in the Pomeron loop  leads to the  value of $b$ of the order the typical size of the dipoles in the loop and can change considerable the conclusions of the approaches where the $b$ dependence has not been taken into account (see \cite{K,STPH,EGM,IMS}).

 For our kernel, that sums log contributions,  the impact parameter dependence does not influence on such summation since the BFKL Pomeron shows the rapid fall down as a function $b$ ($ \propto 1/b^4$) and integration over $b$ does not generate any logarithmic contribution. The $b$ dependence for  our kernel has been discussed in details  in Refs. \cite{GLR,LT} and we will show how it works in the next section.

It should be mention that the relation between fields in momentum and coordinate representation looks as follows
\beq \label{MR}
\Phi(x_1,x_2;Y)\,\,=\,\,x^2_{12}\,\int\,d^2\,k\,e^{i\,\vec{k} \cdot \vec{x}_{12}}\,\,\Phi\left(k,b;Y\right)
\eeq
where $ x_{12}\,=\,x_1 - x_2 $.

We would like to solve directly \eq{ZEQMR}.  Our main assumption is that $Z\left(b,k;Y\right)$ actually is a function of one variable $Z\left(\xi;Y,[u]\right)$
\footnote{For simplicity we neglect the dependence of $Z$ on  impact parameters. For large nuclei this is a good approximation.}:
\beq \label{XI}
\xi\,\,=\,\,\ln\left(Q^2_s(x)/k^2\right)
\eeq

 In finding the solution we replace  the generating functional of \eq{ZMR}
by the generating function
\beq \label{GF}
Z\left(\xi,u \right)\,\,\,=\,\,\,\sum_{n=1}\,\,P_n \left(\xi \right)\,u^n
\eeq

We can use \eq{GF} instead of the generating functional of \eq{ZMR} since the scattering amplitude  is determined by
the following equation \cite{K,L2}
\bea \label{N}
N \Lb Y;k;[\gamma_i] \Rb  &= &- \sum^{\infty}_{n =1}
\,\frac{(-1)^n}{n!}\,
\,\gamma_n(\{k_i\};b ;Y_0;R) \,\,\,\prod^n_{i=1} d^2 k_i \,\,\frac{d^n}{d^n
u}Z\Lb Y-Y_0; \{k_i\},u \Rb|_{u =1}\,\,\rightarrow \nonumber \\
  &\xrightarrow{\xi \,\gg\,1}&- \sum^{\infty}_{n =1}\,\,\,
\,\frac{(-1)^n}{n!}\,\gamma_n(\{k_i=1/R\};b;Y_0;R)\,\,Z\Lb Y-Y_0; \{k_i=1/R^2\},u \Rb|_{u =1}\,\,= \nonumber \\
 &=&- \sum^{\infty}_{n =1}\,\,\,
\,\frac{(-1)^n}{n!}\,\gamma_n(\{k_i =1/R\};b;Y_0;R)\,\,Z\Lb \xi; u \Rb|_{u =1}\,
\eea
$\gamma_n(\{k_i\};b ;Y_0;R)$ in \eq{N}  is the scattering amplitude of $n$ dipoles with momenta $k_i$ with the target of the size $R$ at low energy. In logarithmic approximation we can consider this amplitude as
$\gamma_n(\{k_i\};b ;Y_0;R) = \gamma_n(\{k_i=1/R\};b;Y_0;R)\,\prod^n_i\,\delta^{(2)} (k_i -1/R)$.
Therefore we need to know only generating function of \eq{GF} to find the scattering amplitude.

In the perturbative QCD region ($ r^2\,Q^2_s(x)\,\,\approx\,\,Q^2_s/k'^2\,\,\ll\,\,1$) the scattering amplitude has the following form for the kernel of \eq{CHIM}
\beq \label{DLA}
N\left(Y\,=\,\ln(1/x);\ln(k'^2\,R^2)\right)\,\,=\,\,\as^2\,\exp \left(\,2\,\sqrt{\bas \,Y\,\ln(k'^2\,R^2)}  \,\,\,-\,\,\,\ln(k'^2\,R^2)\,\right)
\eeq
which can be translated in the initial condition, namely,

\beq \label{ICDLA}
Z\left(Y;\ln(k'^2\,R^2); \,u \right)\,\,=\,\,1\,\,\,-\,(1 \,- \,u)\,\,\exp \left(\,2\,\sqrt{\bas \,Y\,\ln(k'^2\,R^2)} \,\,\,-\,\,\,\ln(k'^2\,R^2)\right)
\eeq
In Ref. \cite{IIM} it is proved that
the solution of the linear evolution equation in the vicinity of  $k^2 \to Q^2_s$ shows the geometrical scaling behaviour
 and in terms of the generating function this contribution looks as
\beq \label{XIIC}
Z\Lb \xi; u\Rb |_{\xi < 0, \,\,|\xi| <1}\,\,=\,\,1\,\,-\,\,(1 -u)\,\exp\left(\frac{1}{2}\,\xi \right)
\eeq
for the solution of \eq{ICDLA}. In Ref. \cite{IIM} it is shown that the dependence of the perturbative solution on the only one variable $\xi$  is a general property also for the exact kernel of the BFKL equation.
\eq{XIIC} will be used as the initial condition in our solution.

To check the strategy of our approach, especially \eq{GF} ,  we first consider the BFKL Pomeron interaction in the kinematic region near to the saturation scale where we expect the geometrical scaling solution of \eq{XIIC}.  We will show that in leading log approximation of perturbative QCD which corresponds to the kernel of \eq{CHIM}, we can avoid the assumptions
given by \eq{L12} and \eq{B}, while  \eq{GF} will be proved.

\section{The BFKL Calculus in the geometrical scaling region}
We start with the simplest enhanced diagram of \fig{enhdi}-a.  We will write the expression for this diagram for function $\phi$ which is  defined as
\beq \label{enh1}
\phi(\xi;\rho,q)\,\,\,=\,\,\,\int\,d^2\,b\,\,e^{i\,\vec{q}\cdot\vec{b}}\,\,N\Lb \xi ,\rho;b \Rb
\eeq
where $\xi$ is defined by \eq{XI} and $\rho \,\,=\,\,\ln(k^2\,R^2)$.
The diagram can be written as follows
\beq \label{enh2}
\phi\Lb \mbox{\fig{enhdi}-a} | \xi,\rho;q \Rb\,\,=
\,\,\int\,d\,\xi' d\,\rho' d^2\,q'\,\,d\,\xi" d\,\rho"\,\,\phi\Lb \xi - \xi', \rho - \rho', q \Rb\,
\eeq
$$
G_{3P}\,,\phi\Lb \xi' - \xi", \rho' - \rho", \vec{q} - \vec{q}' \Rb\,\,\phi\Lb \xi' - \xi", \rho' - \rho",  \vec{q}' \Rb\,\,
G_{3P}\,\,\phi\Lb \xi" - \xi_0, \rho" - \rho_0, q \Rb \,
$$

In \eq{enh2} we introduced the variables $\xi$ instead of rapidity $y$ and $\rho $ instead of $k^2$.

The integration over $q'$ in \eq{enh2} has been discussed in details in Refs. \cite{GLR,LW}. The result of this discussion is the fact that the integration over $q'$ sits at $q' = $ the lowest virtuality in the loop. In other words,  $\int\,d^2\,q' \,\,=\,\,\pi\,k"^2 $ in \fig{enhdi}-a if we calculate this diagram in the region of $\xi < 0$. Taking this into account we reduce \eq{enh2} to the following one
\beq \label{enh3}
\phi\Lb \mbox{\fig{enhdi}-a} | \xi,\rho;q =0 \Rb\,\,=\,\,\frac{\as^2}{16}\,\,\int^{\xi}\,\,\int^{\rho}\,d\,\xi' d\,\rho'\,\int^{\xi'}\,\,\int^{\rho'}\,d\,\xi" d\,\rho"\,\,\phi\Lb \xi - \xi', \rho - \rho', q=0 \Rb\,
\eeq
$$
\,\,\phi\Lb \xi' - \xi", \rho' - \rho", q=0 \Rb\,\,\phi\Lb \xi' - \xi", \rho' - \rho",  q=0  \Rb\,\,
\,\,\phi\Lb \xi" - \xi_0, \rho" - \rho_0, q=0 \Rb
$$

It is clear the summation all diagrams can be done using the following functional integral
\beq \label{DLAFI}
Z[\Phi,\Phi^+]\,\,=\,\,\int\,
\,D\,\Phi(\xi,\rho)\,D\,\Phi^+(\xi,\rho)\,e^{S[\Phi,\Phi^+]}
\eeq
with
\beq  \label{DLAMR}
S\,\,=\,\,S_0\,+\,S_I\,\,=
\,\,\int\,d \xi\,d \rho\,\Phi^+\left(\xi,\rho \right)\,\,
\left(4\, \frac{\partial \Phi\left(\xi,\rho\right)}{\,\partial\,\xi}\,\,-\,\,\int^{\rho}\,d\,\rho'\,\,\left (\,4 \frac{\partial \Phi\left(\xi,\rho'\right)}{\,\partial\,\xi}\,\,+\,\,\Phi\left(\xi,\rho'\right)\,\,\,+\right.\right.
\eeq
$$
 \left. \left.
\,\,+ \,\,\,\Phi^+\left(\xi,\rho' \right)\,\Phi\left(\xi,\rho'\right)
+\as^2  \,\left\{\Phi\left(\xi,\rho' \right)\,\Phi^+\left(\xi,\rho'\right)\,\,-\,\,\Phi^+\left(\xi,\rho'\right)\,\Phi\left(\xi,\rho'\right)\,\Phi\left(\xi,\rho'
\right)
\right\}\,\right) \right)
$$

As was shown in Ref.\cite{KLP} this functional integral can be rewritten as the equation for the generating functional , defined as follows
\beq \label{DLAZ}
Z\Lb \xi,[u(\rho)]\Rb\,\,\equiv\,\sum_{n=1}\,\int\,P_n\Lb\xi, \rho_1; \dots; \rho_i;\dots; \rho_n \Rb \,\prod^n_{i=1}
u(\rho_i)\,d\,\rho_i
\eeq
namely,
\beq \label{DLAZEQ}
4\,\frac{\partial\,Z\Lb \xi,[u]\Rb}{\partial \xi}\,=
\eeq
$$
\,\int\,d\,\rho \,\,\int^\rho\,d\,\rho'\,\Lb \frac{\delta \partial\,Z\Lb \xi,[u]\Rb}{\delta u[\rho']\,\partial \xi}
\,-\,\Lb u(\rho)  - u(\rho)\,u(\rho') \,\Rb\,\frac{\delta \,Z\Lb \xi,[u]\Rb}{\delta u[\rho']}\,\,-\,\,\as^2\,\Lb u(\rho)\,u(\rho)  - u(\rho)\Rb \frac{\delta^2\,Z\Lb \xi,[u]\Rb}{\delta u(\rho)\,\delta u(\rho')}\Rb
$$
This functional equation gives the full description in double log approximation  for $\xi < 0$. In particular, it leads to the linear equation for the functional $ N = 1 - Z\Lb \xi,u \,=\,1 - \gamma \Rb$ if we neglect the terms of the order $N^2$.
This equation is
\beq \label{DLALEQ}
4\,\Lb \frac{\partial^2 \,N}{\partial \xi\,\partial\,\rho}\,\,+\,\, \frac{\partial \,N}{\partial \xi}\Rb\,\,=\,\,N
\eeq
It is easy to check that  $N$ of \eq{DLA} satisfies this equation.

However, \eq{DLAZEQ} can be simplified in the region of $|\xi| \,\gg\,1 $ and $\rho\,\gg\,1$  in the kinematic region where
 we have the geometrical scaling solution of \eq{XIIC}. In this region \eq{enh3} has a simpler form, namely,

\bea
\phi\Lb \mbox{\fig{enhdi}-a}| \xi \Rb\,\,&=&~~~~~~~~~~~~~~~~~~~~~~~~~~~~~~~~~~~~~~~~~~~~~~~~~~~~~~~~~~~~~~~~~~~~~~~~~~~~~~~~~~~~~~~~~~~~ \nonumber
\eea
\bea
&=&\,\,\frac{\as^2}{16}\,\,\int^{\xi}_{\xi_0}\,\,d\,\xi'\,\int^{\xi'}_0\, d\,\rho' \,
\int^{\xi'}_{\xi_0}\,\,
d\,\xi"\,\int^{\xi"}_0 d\,\rho"\,\,
\,\,\phi\Lb \xi - \xi' \Rb\,
\,\,\phi\Lb \xi' - \xi" \Rb\,\,\phi\Lb \xi' - \xi" \Rb\,\,
\,\,\phi\Lb \xi" - \xi_0\Rb \label{enh4} \\
 &=&\frac{\as^2}{16}\,\,\,\,\,\int^{\xi}_{\xi_0}\,\,\xi'\,d\,\xi'\, \,\int^{\xi'}_{\xi_0}\,\,\xi"\,d\,\xi"\,\,
d\,\xi"\,\int^{\xi"}_0 d\,\rho"\,\,
\,\,\phi\Lb \xi - \xi' \Rb\,
\,\,\phi\Lb \xi' - \xi" \Rb\,\,\phi\Lb \xi' - \xi" \Rb\,\,
\,\,\phi\Lb \xi" - \xi_0\Rb \label{enh5}
\eea

\FIGURE{
\begin{tabular}{c c}
\epsfig{file=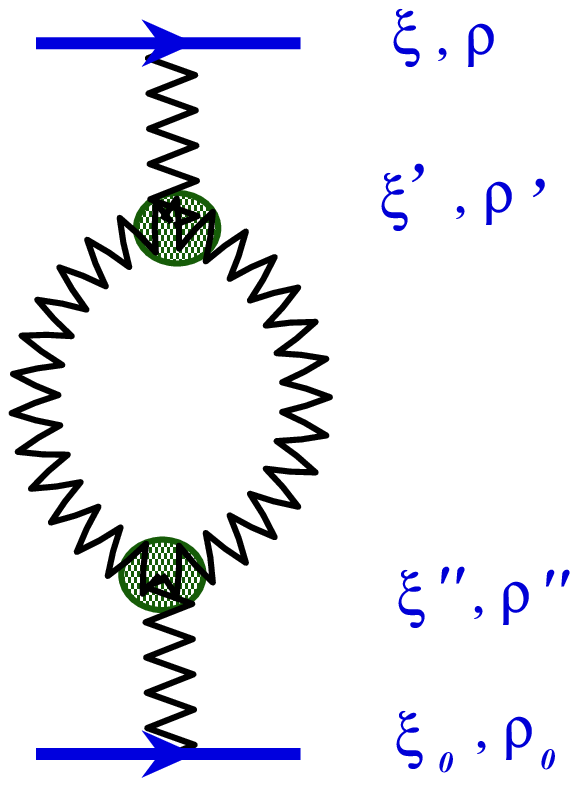,width=85mm,height=70mm} & \epsfig{file=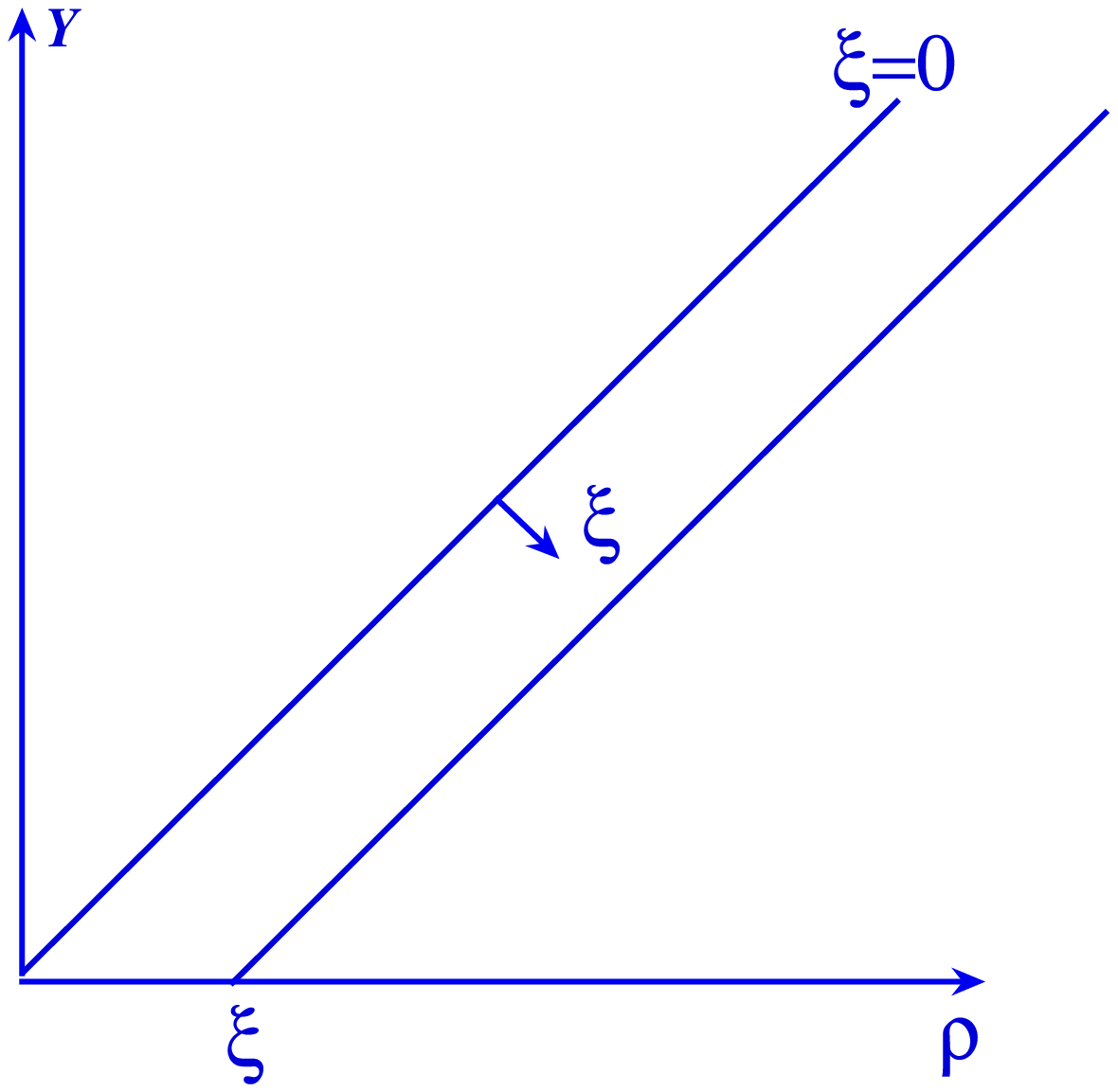,width=75mm} \\
\fig{enhdi}-a & \fig{enhdi}-b \\
\end{tabular}
\caption{ The simplest enhanced diagram for interaction of the BFKL Pomeron (\fig{enhdi}-a) and
the plot  (\fig{enhdi}-b) for integration variables:  \,$\xi = 4\,\as\,Y\,\,-\,\,\rho$
\,\,with $\rho\,\,=\,\,\ln (k^2\,R^2)$.  $Y$  is rapidity. }
\label{enhdi}
}

In calculating  \eq{enh4} we restrict ourselves by the kinematic region $\xi <0$
where we can safely use \eq{XIIC} for the BFKL Pomeron.
 All integrations in
  \eq{enh4} and \eq{enh5}  are taken in the so called double log approximation
 which corresponds to \eq{CHIM} for the kernel and \eq{ICDLA}
 for the solution of the linear equation. We also used the kinematic restriction $\rho < \xi$   that follows from \fig{enhdi}-b.

\eq{enh5} suggests that the problem of summing all BFKL Pomeron diagrams in the saturation region for $\xi \gg 1$  can be reduced to the solution  of the following equation for the generating function defined by \eq{GF} and \eq{N}
\beq \label{GFEQG}
4\,\frac{\partial Z\Lb  \xi; u\,\Rb}{\partial \xi}\,\,\,=\,\,\,-\,\xi\,u\,(1 - u)\,\,\frac{\partial Z\Lb  \xi; u\,\Rb}{\partial u}
\,\,\,\,+\,\,\,\,\as^2\,\,\xi\,\,u\,(1 - u)\,\frac{\partial^2 Z\Lb  \xi; u\,\Rb}{\partial u^2}
\eeq
This equation can be rewritten in more convenient form introducing   \,\, ${\cal Y} \,=\,\xi^2/8$,\,\, namely,
\beq \label{GFEQG1}
\frac{\partial Z\Lb  {\cal Y}; u\,\Rb}{\partial {\cal Y}}\,\,\,=\,\,\,-\,u\,(1 - u)\,\,\frac{\partial Z\Lb  {\cal Y}; u\,\Rb}{\partial u}
\,\,\,\,+\,\,\,\,\,\,\as^2\,u\,(1 - u)\,\frac{\partial^2 Z\Lb  {\cal Y}; u\,\Rb}{\partial u^2}
\eeq

\eq{GFEQG1} is the same as for the BFKL Pomeron calculus in zero transverse dimension (see Ref. \cite{KOLE1} and references therein). This equation has been solved \cite{KOLE1} .  We will discuss below the main propertiers of the solution to \eq{GFEQG1} but first we start  from the solution to our problem in mean field approximation to check the main assumption that we make on the way of obtaining \eq{GFEQG1}.

\section{Exact solution and the mean field approximation (MFA)}
In the mean field approximation $\chi[u]$ is determined only by \eq{chimr} . This fact considerably simplifies the
equation since such an equation belongs to the well known class of the Louisville equations. The  Louisville equation has a general solution, namely, $Z\left(Y - F(u) \right)$. Using this form for the general solution as well as two conditions (initial and boundary, see \eq{IC} and \eq{BC}):
we can easily prove that for $Z$ we have the non-linear equation \cite{MUCD,L1}, namely
\beq \label{NLEQ}
\frac{\partial \,Z\Lb Y-Y_0; b,k\,\Rb}{
\partial \,Y}\,\,=\,\,  \frac{\bas}{2 \pi}\,\left( - \int\,K(k,k')\,Z\Lb Y-Y_0; b,k'\,\Rb\,\,+\,\,Z^2\,\Lb Y-Y_0; b,k\,\Rb\right)
\eeq
This equation leads to Balitsky-Kovchegov equation\cite{K,B}  for the scattering amplitude and it has been solved in Ref. \cite{LT} and the solution turns out to be the function of the only one variable $\xi$ (see \eq{XI}).
In Ref. \cite{LT} it is shown that at large $\xi$ the generating functional $Z\Lb Y-Y_0; b,k'\,\Rb$  for the kernel of \eq{CHIM}  behaves as
\beq \label{LTSOL}
Z\Lb Y-Y_0; b,k'\,\Rb\,\,\,=\,\,C\,\exp \left(-\frac{ \xi^2}{8} \right)
\eeq
In  the log  approximation for the BFKL kernel in the saturation region which is related to the kernel of \eq{CHIM},  (see Ref. \cite{LT} and references therein) we have :
\beq \label{APRKER}
\int\,d^2\,k'\,\,K\left(k,k'\right)\,\,Z\left(b,k';Y\right)\,\,\,\rightarrow\,\,\int^{Q^2_s(x)}_{k^2}\,\,\frac{d k^2}{k^2}
\,\,Z\left(b,k';Y\right)
\eeq

In the MFA \eq{GFEQG1} reduces to very simple Louisville    equation, namely
\beq \label{MFA1}
\frac{\partial Z\Lb  {\cal Y}; u\,\Rb}{\partial {\cal Y}}\,\,\,=\,\,\,-\,u\,(1 - u)\,\,\frac{\partial Z\Lb  {\cal Y}; u\,\Rb}{\partial u}
\eeq
which has the solution
\beq \label{MFA2}
Z\Lb  {\cal Y}; u\,\Rb\,\,\,=\,\,\,\frac{u}{1 \,\,\,+\,\,\,(1 - u)\,\Lb \exp\Lb  {\cal Y}\Rb\,\,-\,\,1 \Rb}
\eeq

One can see that \eq{MFA2} satisfies the initial condition of \eq{XIIC} if we  define ${\cal Y} \,\,=\,\,\xi^2/8\,\,-\,\xi/2$.
For $\xi \,\to\,0$ the scattering amplitude is equal to
\beq \label{MFA3}
N\Lb {\cal Y},\gamma= 1 - u \Rb\,\,\,=\,\,\,\frac{\gamma\,\, \exp\Lb  {\cal Y}\Rb}{1 \,\,\,+\,\,\,\gamma\,\Lb \exp\Lb  {\cal Y}\Rb\,\,-\,\,1 \Rb}
\eeq
Since $\gamma $ is of the order of $\as$ we obtain \eq{DLA} at small $\xi$ with our definition of ${\cal Y}$.
Of course, we cannot guarantee term $\frac{1}{2}\xi $ since we assumed that $\xi \,\gg\,1$, therefore,   we add
this term
to guarantee the
matching with the initial conditions at $\xi \to 0$.

Therefore, we showed that our procedure involving \eq{GF} and \eq{XIIC} leads to the correct behaviour of the amplitude in the mean field approximation.  Therefore,  we confirm our  main assumption that for searching for behaviour at large values of $\xi$,  we can consider of generating function of \eq{GF} instead of the generating functional of \eq{ZMR}.

\section{A possible  solution with the geometrical scaling behaviour}
\subsection{General properties}
\eq{GFEQG1} has been solved in Ref. \cite{KOLE1}., where it shown that the solution for the scattering amplitude approaches  at high energy behaves as
\beq \label{SCSOL1}
N\Lb \xi \Rb\,\,\,\,\rightarrow\,\,1\,\,-\,\,\exp \Lb - C(\kappa)\,\xi^2/2 \Rb
\eeq Ref.
where $\kappa = 1/\as^2$. The exact value of the factor $C(\kappa)$ is found in  Ref. \cite{KOLE1}, however, here we will illustrate the main  property of the exact solution using an approximate semi-classical  method developed in Ref. \cite{L4}
which explores the large value of $\kappa$.

First we find the asymptotic solution to the master equation from the following equation:
\beq  \label{GEQGF}
0\,=\,\, - \,u \,(\,\,-\,\,u) \,\,
\frac{\partial Z\Lb \xi; u\Rb}{\partial u}\,\,\,+\,\,\,\frac{1}{\kappa}\,u\,( 1 \,\,-\,\,u)\,
\frac{\partial^2 Z\,\Lb \xi;u\Rb}{\partial u^2}
\eeq
The solution that satisfies the boundary condition of \eq{BC} is equal to
\beq \label{SCSOL2}
Z_{asymp}\Lb {\cal Y} = \infty,u \Rb\,\,\,=\,\,\,\frac{ 1 \,\,-\,\,e^{\kappa\,u}}{1\,\,\,-\,\,\,e^{\kappa} }
\eeq

The second step is to search the solution in the form
\beq \label{SCSOL3}
 Z\Lb \xi; u\Rb\,\,=\,\,Z_{asymp}\Lb {\cal Y} = \infty,u \Rb\,\,\Delta Z\Lb \xi; u\Rb
 \eeq
 assuming that $ \partial^2 \Delta Z\Lb {\cal Y}; u\Rb/\partial u^2\,\,\ll\,\,\kappa\,\,\Lb,\partial  Z_{asymp}\Lb {\cal Y} = \infty,u \Rb/\partial u \,\Rb\,\Lb \partial \Delta Z\Lb {\cal Y}; u\Rb/\partial u \Rb$. Since $\kappa \,\gg\,1$ this assumption looks reasonable. For $\Delta Z\Lb \xi; u\Rb$ we obtain the equation
 \beq \label{SCSOL4}
\frac{\partial \Delta Z\,\Lb {\cal Y} u\Rb}{\partial \,{\cal Y}}\,\,\,= \,\,\,-\, u\,(1 - u)
\frac{\partial \Delta Z\,\Lb {\cal Y} u\Rb}{\partial u}\,\,\,+\,\,2\,u (1 - u)\,\frac{\partial Z_{asymp}\,\Lb {\cal Y}=\infty,  u\Rb}{\partial u}
\frac{\partial \Delta Z\,\Lb {\cal Y} u\Rb}{\partial u}
\eeq

Assuming that $\kappa \,\gg\,1$  \eq{SCSOL4} reduces to the form
\beq \label{SCSOL5}
\frac{\partial \Delta Z\,\Lb {\cal Y}; u\Rb}{\partial \,{\cal Y}}\,\,=\,\,\,u\,(1 - u)\, \frac{\partial \,\Delta Z\,\Lb {\cal Y}; u\Rb}{\partial u}
\eeq
\eq{SCSOL5} has the solution
\beq \label{SCSOL6}
\Delta Z\,\Lb {\cal Y}; u\Rb\,\,\,=\,\,\Phi \Lb \,{\cal Y} \,+\,\ln (\frac{u}{1 - u} \Rb
\eeq
with function $\Phi$ that should be found from the initial conditions of \eq{IC}.

The solution is
\beq \label{SCSOL7}
Z_{asymp}\Lb {\cal Y}=\infty;  u \Rb\, \Delta Z\,\Lb {\cal Y}; u\Rb\,\,\,=\,\,\,\tilde{u} \frac{1\, -\, e^{\kappa\,u}}{1 \,-\,e^{\kappa\,\tilde{u}}}
\,\,\,\,\,\mbox{with}\,\,\,\,\,\,\tilde{u}\,\,\,=\,\,\,\frac{u}{u\,\,+\,\,(1 -u)\,\exp\Lb - {\cal Y} \Rb}
\eeq
However we cannot use directly \eq{SCSOL7} to determine the asymptotic behaviour of the scattering amplitude at high ${\cal Y}$ since we need to find the typical values of $u$ for such an amplitude at large values of $\xi$ or
to find the solution in the entire kinematic region. The latter we cannot do since our equation is valid only at $\xi \,\gg\,1$.
To solve this problem at $\xi\,\gg\,1$ we need to use the $t$=channel unitarity constraints ( see Refs. \cite{IM,KOLEIM,L4}).
we need to use the $t$-channel unitarity constraint ( see Refs. \cite{IM,KOLEIM,L4}), which has the following form
\beq \label{SCSOL71}
N^{MPSI}\Lb  \xi \Rb \,\,\,= \,\,\sum_{n=1}^{\infty}\,\,\frac{(-1)^n}{\,n!\,}\,\,
\frac{\delta^n Z\Lb  \xi\,-\,\xi'; u\Rb}{\delta u^n}|_{u=1}\,\frac{\delta^n Z\Lb   \xi'; v\Rb}{\delta v^n}|_{v =1}
\,\gamma^{BA}
\eeq
where $\gamma^{BA}$ is the amplitude in the Born approximation for the interaction of two dipoles at low energy, which is equal to $\as^2 = 1/\kappa$ in our case.
Using $\kappa\,\gg\,1$ we see that in this limit $\frac{d^n Z}{d u^n} = \Lb \kappa \frac{d \tilde{u}}{d u}\Rb^n Z$
and $ \frac{d \tilde{u}}{d u}\,=\,\exp\Lb - {\cal Y} \Rb$ at $u =1$.

Using these observation we obtain that
\bea \label{SCSOL72}
N^{MPSI}\Lb  \xi \Rb \,\,\,&\xrightarrow{\xi\,\gg\,1}&\,\,1\,\,-\,\,\exp \left\{ -\,\as^2\,\kappa^2\,\Lb\,1\,\,-\,\,e^{ - \frac{(\xi - \xi')^2}{8}}\,\Rb\,\Lb\,1\, -\,\,e^{-\, \frac{ \xi'^2}{8}}\Rb \right\}\,\,\nonumber \\
&\xrightarrow{\xi\,\gg\,1, \xi' = 1/2 \xi}&\,\,1 \,-\,e^{-\as^2 \kappa^2}\,\,-\,\,2\kappa\,\,e^{- \xi^2/32}
\eea
The choice $\xi' = \xi/2$ is clear since it gives the slowest  approach to $N =1$. \eq{SCSOL72} can only  be considered
 as a approximation to  the exact solution of Ref.\cite{KOLE1} .
It should be stressed that \eq{SCSOL72} leads to quite a  diffrent result in comparison with the Mueller-Patel-Salam-Iancu
approach \cite{IM}.
However, the  MPSI approach has an advantage to be correct also for $\xi \approx 1$. This is the reason why we consider our problem in this approach in the next section.
\subsection{The Mueller-Patel-Salam-Iancu (MPSI) approach}
 This approximation has  a  legitimate region of applicability \cite{IM,KOLEIM}; can be used for small values of $\xi$  and
it describes quite well in the entire kinematic region the exact solution to the BFKL Pomeron calculus
in zero transverse dimension \cite{KLLM}.

In this approach the scattering amplitude can be calculated using the following formula \cite{IM,KOLEIM}
\bea \label{SCSOL8}
&&N^{MPSI}\Lb x,y; x',y'; Y \Rb \,\,\,= \\
&&\,\,\sum_{n=1}^{\infty}\,\,(-1)^n\,n!\,\int \prod^n_{i=1}\,d\,x_i\,d\,y_i\,d\, x'_i d\,y_i
\frac{\delta^n Z^{MFA}\Lb  Y - Y'; [u_i]\Rb}{\delta u^n_i}|_{u_i=1}\,\frac{\delta^n Z^{MFA}\Lb   Y'; [v_i]\Rb}{\delta v^n_i}|_{v_i=1}
\,\gamma^{BA}(x_i,y_i; x'_i, y'_i) \nonumber
\eea
where $\gamma^{BA}(x_i,y_i; x'_i, y'_i)$ is the amplitude of interaction of two dipoles in the Born approximation of perturbative QCD.

In the case of the geometrical scaling solution the MPSI approximation formula has an elegant form \cite{KOVGL}, namely
\beq \label{SCSOL9}
N^{MPSI}\Lb \xi \Rb \,\,\,=\,\,\,1\,\,\,-\,\,\left\{\exp\Lb - \gamma\,\frac{d}{d u}\,\frac{d}{d v}\Rb\,N^{MFA}\Lb \xi - \xi';u \Rb
\,N^{MFA}\Lb  \xi' ,v\Rb\,\right\}|_{u=1; v = 1}
\eeq
where the generating function  $\,N^{MFA}\Lb  \xi ,u\Rb$ is  defined as
\beq \label{SCSOL10}
N^{MFA}\Lb  \xi ,u\Rb\,\,\,=\,\,1\,\,-\,\,Z^{MFA}\Lb \xi,u \Rb
\eeq

The generating functional in MFA for the kernel of \eq{CHIM} has been found in Ref. \cite{LT}. In this paper it is shown that
the solution to \eq{NLEQ} has the form that in our notations looks as follows
\beq \label{MPSI1}
N\Lb \xi, \gamma  \equiv  1 - u \Rb \,\,\,=\,\,\,1\,\,\,-\,\,\exp\Lb - \zeta(\xi,u) \Rb
\eeq
where function $\zeta(\xi,u)$ can be calculated from the implicit equation
\beq \label{MPSI2}
\xi\,\,=\,\,\sqrt{2}\,\int^{\zeta}_{\gamma}\,\,\frac{d\,\zeta'}{\sqrt{\zeta'\,\,+\,\,\Lb \exp(- \zeta')\,-\,1 \Rb}}
\eeq
For small values of $\zeta \to \gamma$ \eq{MPSI2} gives
\beq \label{MPSI3}
\ln\Lb\zeta(\xi,u)/\gamma \Rb\,\,\,=\,\,\frac{1}{2} \,\xi\,\,\,\,\mbox{which leads to}\,\,\,\,\zeta(\xi,u)\,\,=\,\,\gamma  e^{\frac{\xi}{2}}\,\,\,\xrightarrow{\xi\,\ll\,1}\,\,\,\gamma\,\Lb 1 + \frac{\xi}{2} \Rb
\eeq
It is easy to see that \eq{MPSI1} with $\zeta(\xi,u)$, determined by \eq{MPSI3}, satisfies the initial conditions of \eq{XIIC},
namely,
\beq \label{MPSI4}
N\Lb \xi = 0, \gamma\Rb\,\,=\,\,\gamma\,\,\,\,\,\,\,\,\,\,\,\,\mbox{and}\,\,\,\,\,\,\,\,\,\,\,\,\frac{d\,\ln N\Lb \xi, \gamma\Rb}{d\,\xi}|_{\xi =0}\,\,=\,\,\frac{1}{2}
\eeq
For large $\zeta$ \eq{MPSI2} leads to
\beq \label{MPSI5}
\zeta(\xi,u)\,\,=\,\,\frac{\xi^2}{8}
\eeq
 which does not depend on $u$.  Using this limiting behaviour we can suggest the following extrapolation formula
 \bea
 \xi\,\,&=&\,\,\,\,\sqrt{2}\,\int^{\zeta}_{\gamma}\,\,d\,\zeta' \Lb\frac{1}{\sqrt{\zeta'\,\,+\,\,\Lb \exp(- \zeta')\,-\,1 \Rb}}
\,\,-\,\,\frac{\sqrt{2}}{\zeta'}\Rb \,\,+\,\,2\,\ln(\zeta/\gamma) \label{MPSI61}\\
&=&\,2\,\sqrt{2\,\zeta}\,\,\,+\,\,\,2\,\ln(\zeta/\gamma) \label{MPSI62}
\eea
In the first integral in \eq{MPSI61} the small values of $\zeta'$ does not contribute and we can replace it by the asymptotic al behaviour at large $\zeta'$.

Using \eq{MPSI3},\eq{MPSI5} and \eq{MPSI62} we can suggest an approximation for the solution of \eq{MPSI2}, namely.
\beq \label{MPSI9}
\zeta\Lb \xi; \gamma \Rb\,\,\,=\,\,\,\gamma (1 + \xi/2) \,\,+\,\,\xi^2/8
\eeq

Substituting \eq{MPSI9} into \eq{SCSOL9} and choosing $\xi' = \xi/2$  that correspond to the smallest corrections (see \cite{IM} ) , we obtain the final result
\beq \label{MPSI10}
N^{MPSI}\Lb \xi \Rb\,\,\,=\,\,\,1\,\,\,\,-\,\,\,\exp \Lb -\,\xi^2/16\,\,-\gamma\,(1 \,+ \,\xi/2) \Rb
\eeq
Therefore, the MPSI approximation leads to the geometrical scaling which satisfies   the initial conditions at $\xi =0$. The $\xi^2$ term does not depend of the scattering amplitude $\gamma \propto \as^2$ while the next terms show such dependence. Comparing with the solution of \eq{SCSOL72} one can see that both have the same $\xi^2$ dependence but the coefficients in front of of this term is quite different. We need to recall that the MPSI approach cannot be correct at very high energies \cite{IM}.

\section{Conclusions}
The main result of this paper is the reduction of the complicated problem of summing all Pomern loops to a  Pomeron calculus in the zero transverse dimension but for 'time'  ${\cal Y}$ given by ${\cal Y} = \xi^2/8$ at $\xi \,\gg\,1$. The principle features of this solution is the fact that the scattering amplitude depends on the only one variable: $Q^2_s(x)\,r^2$ where $Q_s(x)$ is the saturation scale while $r$ is the size of the scattering dipole. Such geometrical scaling behaviour
has been proven for the mean field approximation and has been seen experimentally \cite{GS}. However,   this scaling behaviour was  not expected for the theory given by the functional integral of \eq{FI}.  Indeed, at first sight,  such theory can be rewritten as
the Langevin equation for directed percolation \cite{HH}, namely,
\beq \label{DPG}
\frac{\partial \,\,\Phi\left(b,k; Y\right)}{\partial Y}\,\,\,
=\,\,\frac{\bas}{2\pi}\,\int\,d^2\,k'\,K\left(k,k'\right)\,
\Phi\left(b,k';Y\right)\,\,\,-\,\,\,\frac{2\,\pi\,\bas^2}{N_c}\cdot \, \Phi( b,k;Y)\,\Phi(b,k;Y)
\,\,+\,\,\zeta(b,k; Y)
\eeq
with
\beq \label{DPNOISE}
<|\zeta(b,k; Y)|>\,\,=\,\,0\,;
\eeq
$$
\,\,\,\,\,\,\,\,\,\,\,<|\zeta(b,k; Y)\,\zeta(b',k'; Y')
|>\,\,=\,\,\frac{4\,\pi\,\bas^2}{N_c}\cdot \Phi(b,k;Y)\,\left( 1 - \Phi(b,k;Y) \right)\,\,\delta^{(2)}(\vec{b}
-\vec{ b}')\,\,k^2\,\delta^{(2)}(\vec{k} -
\vec{k}')\,\,\delta\left(Y - Y'\right)
$$

As it has been shown in Refs. \cite{STPH,EGM,IMS} we rather expect from the statistical physics analogy a solution of
the following type
\beq \label{SOLSP}
N\Lb \tilde{\xi};Y\Rb\,\,=\,\,\frac{1}{\sqrt{2 \pi}\,\sigma(Y)}\,\int d \ln( Q^2_s\,R^2)\,\, N\Lb  \ln(Q^2_s/k^2) \Rb\,\,\exp \Lb - \frac{
\Lb \ln ( Q^2_s\,R^2)\,\,-\,\,<\ln (Q^2_s\,R^2)> \Rb^2}{2\,\sigma^2(Y)} \Rb
\eeq
where $ N\Lb  \ln(Q^2_s/k^2) \Rb$ is the scaling solution in the mean field approximation;  $<\ln Q^2_s>$ is the average
saturation scale which differs from $Q_s$ given by \eq{QSCHIM} or \eq{QSCOR}, namely,\cite{STPH}
\beq \label{ACQS}
<\ln( Q^2_s\,R^2)>\,\,=\,\,\ln( Q^2_s\,R^2)\,\,\,-\,\,\frac{\pi^2 \bas}{2}\,\frac{\gamma_{cr}\,\chi''(\gamma_{cr})}{\ln^2(1/\as^2)}\,Y
\eeq
and variance of the saturation momentum $\sigma$ is proportional to
\beq \label{VAR}
\sigma^2\,\,=\,\,<\ln^2 ( Q^2_s\,R^2)>\,\,-\,\,(<\ln (Q^2_s\,R^2)>)^2\,\, \propto\,\,\frac{\bas}{\ln^3(1/\as^2)}\,Y
\eeq

The Pomeron loops change drastically the geometrical scaling behaviour of the MFA solution in an obvious  contradiction with our solution.  Indeed, \eq{SOLSP}   means that  the amplitude  is a function
of one variable
\beq \label{NEWSC}
N \Lb \tau \Rb\,\,\,=\,\,\,N\Lb \frac{\ln ( k^2\,R^2)\,\,-\,\,<\ln (Q^2_s(Y)\,R^2)>}{\sqrt{\bas\,Y/\ln^3(1/\as^2)}} \Rb
\eeq

Despites that \eq{NEWSC} looks very plausible we should not forget that it is correct for the functional integral of \eq{FIMR} which does not treat the impact parameter behaviour  in consistent way. As has been demonstrated in section 3
\eq{B}, which is needed for a proof of \eq{FIMR}, cannot be valid in the Pomeron loops.

The reason for such difference  between our approach and \eq{NEWSC} is clear from the consideration in section 3: we found that not all Pomeron loops  lead to a violation of the geometrical scaling behaviour. Some of them contribute to a change of the scaling solution in the MFA.

 In our approach we have the following equation for the directed percolation instead of \eq{DPG}
\beq \label{DPGN}
4\,\frac{\partial \,\,\Phi\left(\xi \right)}{\partial \xi}\,\,\,
=\,\,\xi
\Phi\left(\xi'\right)\,\,\,-\,\,\,\frac{2\,\pi\,\bas^2}{N_c}\cdot \,\xi\, \Phi( \xi)\,\Phi(\xi)
\,\,+\,\,\zeta(\xi)
\eeq
with
 \beq \label{DPNOISEN}
<|int \,\zeta(\xi)|>\,\,=\,\,0\,;
\eeq
$$
\,\,\,\,\,\,\,\,\,\,\,< |\zeta(\xi)\, \zeta(\xi')
|>\,\,=\,\,\frac{4\,\pi\,\bas^2}{N_c}\cdot \,\,\xi\,\Phi(\xi)\,\left( 1 - \Phi(\xi) \right)\,\,\delta(\xi - \xi')
$$

These equations are equivalent to one dimensional Pomeron theory. The first and, perhaps,  the only that this theory does,  it leads to replacement of the MFA solution in \eq{SOLSP} by the solution that has been discussed here.
\eq{DPNOISEN} suggests that the correlation term in \eq{DPNOISE} has a more complicated structure. We propose to sum \eq{DPNOISEN} and \eq{DPNOISE} but , at the moment, we do not  understand the correct $b$ dependence in \eq{DPNOISE}.

We would like to stress that we suggest a solution only for $\xi\,\gg\,1$ and matching with the initial conditions was done in an approximate approach (the MPSI approach). Therefore, in principle, could be that the solution  we advocate here, just a wrong one   but we believe that the MPSI approximation works well enough to be an argument against such a possibility.

On the other hand,
\eq{SOLSP} is proven only in the vicinity  of the saturation scale where, indeed, $N\Lb  \ln(Q^2_s/k^2) \Rb$ is the solution for MFA  which is averaged with a  Gaussian weight. However, it is difficult to understand how \eq{NEWSC} can
be matched with the perturbative QCD solution of \eq{ICDLA} or \eq{XIIC}  outside  the saturation kinematic region.
Indeed, $ \partial\ln  N/\partial \ln(k^2\,R^2) =  (1/\sigma(Y))\,d\ln  N/d \tau\,\,\ll\,1$ instead of  $ \partial\ln  N/\partial \ln(k^2\,R^2)\,\approx\, 1$.

At first sight the numerical calculation in Refs.  \cite{MSX,STPH,EGM,IMS,NCR} can give us an information on how
the solution behaves  deeply inside of the saturation region. However, it is not the case at the moment , since: (i) all calculations have been performed for the diffusion kernel of \eq{bfkldif} (in other words,  for the F-KPP equation) that cannot pretend to describe  approaching to the asymptotic behaviour; (ii) the hope that such calculation can be generalized to the exact BFKL kernel is based on some conditions (see Rev. \cite{NCR}) which are plausible only in vicinity of the saturation scale; (iii)  it  has not been checked that the numerical solution reproduces  the solution of \eq{ICDLA} or \eq{XIIC} , rather from fig. 16 of Rev.\cite{NCR} one can conclude that they do not approach  them; and (iv)
the accuracy of the numerical calculations has not been checked by a comparison
with the exactly solved model of the BFKL Pomeron calculus in zero transverse dimension.  For the BFKL Pomeron calculus in zero transverse dimension the numerical calculation should coinside with the exact solution given in Ref. \cite{KOLE1} with \eq{DPNOISE} and should lead to decreasing amplitude at high energy (see Ref.\cite{MUBO} and references therein) for  \eq{DPNOISE} without
$ \Phi^2(b,k;Y) $ term.

It should be stressed that the experimental data do not show any deviation from the geomerical scaling behaviour while  the estimates with $\as \approx 0.25$ show that we expect a considerable violation of this scaling behaviour accordingly to \eq{SOLSP}.

Summarising  we do not think that there is any proof that our solution is inconsistent but , of course, we need to remember that this solution cannot be justified in the vicinity of the saturation scale. The only that we showed  is that
our solution does not contradict the exact solution in the perturbative QCD region (see \eq{ICDLA} or \eq{XIIC}).
 Such strategy was  successful in the case of the mean field approximation \cite{LT}.    On the other hand,
solution of \eq{SOLSP} is able to describe the system in  the vicinity of the saturation scale but it is not hundred percents clear how this solution could  be matched with \eq{ICDLA} or \eq{XIIC}. We would like to stress again that the most probable way out is that we need to replace the MFA solution by ours in \eq{SOLSP}.

We are working  to include the exact BFKL kernel in our consideration and to understand better the kinematic region of matching with the perturbative QCD solution.  This is a very important  issue since the model kernel of \eq{CHIM} does not give a correct high energy solution in the MFA (see Ref. \cite{KOLE}). If we will use the correct solution in the MPSI approximation we will obtain the solution which being substituted in \eq{SOLSP} leads to a geometrical scaling behaviour after integration over $\ln(Q^2_s\,R^2)$ . The only change is to replace $Q^2_s$ in the definition of variable $\xi$ (see \eq{XI}) by the new saturation momentum of \eq{ACQS}.

 We hope that the paper will be useful  \,\,  in searching a solution in the saturation region beyond the scope of statistical physics analogy for the BFKL Pomeron calculus including the Pomeron loops.

\section*{Acknowledgments:}
We are very grateful to Asher Gotsman, Edmond Iancu, Michael Kozlov, Uri Maor, Jeremy Miller and  Al Mueller
for hot  and useful discussions on the subject.
This research was supported in part  by the Israel Science Foundation,
founded by the Israeli Academy of Science and Humanities and by BSF grant \# 20004019.

\end{document}